\documentclass[a4paper,12pt]{article}                     
\usepackage{graphicx}
\usepackage{amsmath,epsfig,subfigure}
\usepackage[usenames]{color}

\renewcommand{\mathbf}[1]{\boldsymbol{#1}}

\makeatletter
\renewcommand*\env@matrix[1][*\c@MaxMatrixCols c]{%
  \hskip -\arraycolsep
  \let\@ifnextchar\new@ifnextchar
  \array{#1}}
\makeatother

\newcommand{\tens}[1]{\boldsymbol{\sf{#1}}}
\begin{document}

\title{Nonlinear Correction to the  Euler  Buckling \\ Formula  for Compressed Cylinders \\
with Guided-Guided End Conditions}

\author{Riccardo De Pascalis$^a$, Michel Destrade$^b$, Alain Goriely$^c$,\\[12pt]
  $^a$CNRS, UMR 7190, Institut Jean Le Rond d'Alembert,  \\
  F-75005 Paris, France.\\[12pt]
           $^b$School of Electric, Electronic and Mechanical Engineering, \\
           University College Dublin, \\
           Dublin 4, Ireland.\\[12pt]
           $^c$OCCAM, Institute of Mathematics,\\ University of Oxford, UK}

\date{}

\maketitle

\begin{abstract}
Euler's celebrated buckling formula gives the critical load $N$ for the buckling of a slender cylindrical column with radius $B$ and length $L$ as 
\[
N / (\pi^3 B^2) = (E/4)(B/L)^2,
\]
where $E$ is Young's modulus.
Its derivation relies on the assumptions that linear elasticity applies to this problem, and that the slenderness $(B/L)$ is an infinitesimal quantity. 
Here we ask the following question:
What is the first nonlinear correction in the right hand-side of this equation when terms up to $(B/L)^4$ are kept? To answer this question, we specialize the exact solution of incremental non-linear elasticity for the homogeneous compression of a thick compressible cylinder with lubricated ends to the theory of third-order elasticity.  In particular, we highlight the way second- and third-order constants ---including Poisson's ratio--- all appear in the coefficient of $(B/L)^4$.

\end{abstract}


\newpage

\section{Introduction}
\label{intro}


One of the first, and most important, problems to be tackled by the theory of linear elasticity is that of the buckling of a column under an axial load. 
Using Bernoulli's beam equations, Euler found the critical load of compression $N_\text{cr}$ leading to the buckling of a slender cylindrical column of radius $B$ and length $L$.
As recalled by Timoshenko and Gere \cite{tige61}, Euler looked at the case of an ideal column with (built in)-(free) end conditions. 
What is now called ``Euler buckling'', or the ``\emph{fundamental case} of buckling of a prismatic bar'' \cite{tige61} corresponds to the case of a bar with hinged-hinged end conditions. 
The corresponding critical load is given by
\begin{equation} \label{euler}
\dfrac{N_\text{cr}}{\pi^3B^2} = \frac{E}{4}\left(\frac{B}{L}\right)^2,
\end{equation}
where $E$ is the Young modulus.
The extension of this formula to the case of a thick column is a non-trivial, even sophisticated, problem of non-linear three-dimensional elasticity. 
In general, progress can be only made by using reductive (rod, shells, etc.) theories. 
However, there is another choice of boundary conditions for which the criterion~(\ref{euler}) is valid: namely, the case where both ends are ``guided'' or ''sliding'' (the difference between the two cases lies in the shape of the deflected bar, which is according to the half-period of a sine in one case and of a cosine in the other case.) 
In exact non-linear elasticity, there exists a remarkable three-dimensional analytical solution to this problem (due to Wilkes \cite{Wilk55}) which describes a small-amplitude (incremental) deflection superimposed upon the large homogeneous deformation of a cylinder compressed between two lubricated platens. 
In this case, the Euler formula can be extended to the case of a column with finite dimensions, for arbitrary constitutive law.

The exact incremental solution allows for an explicit derivation of Euler's formula at the \emph{onset of non-linearity}, which combines third-order elastic constants with a term in $(B/L)^4$.
In Goriely et al. \cite{Goriely08}, we showed that for an  incompressible cylinder, 
\begin{equation}\label{incompressible_case}
\frac{N_\text{cr}}{\pi^3B^2} = \frac{E}{4}\left(\frac{B}{L}\right)^2
 - \frac{\pi^2}{96}\left(\frac{20}{3}E + 9 \mathcal{A}\right)\left(\frac{B}{L}\right)^4,
\end{equation}
where $\mathcal{A}$ is Landau's third-order elasticity constant.
This formula clearly shows that \emph{geometrical non-linearities} (term in $(B/L)^4$) are intrinsically coupled to \emph{physical non-linearities} (term in $\mathcal{A}$) for this problem.
(For connections between Euler's theory of buckling and incremental incompressible nonlinear elasticity, see the early works of Wilkes \cite{Wilk55}, Biot \cite{Biot63}, Fosdick and Shield \cite{FoSh63}, and the references collected in  \cite{Goriely08}.)

Now, in third-order \emph{in}compressible elasticity, there are two elastic constants \cite{Ogden74b}: the shear modulus $\mu$ ($=E/3$) and $\mathcal{A}$.
In third-order \emph{compressible} elasticity, there are five elastic constants:
$\lambda$ and $\mu$, the (second-order) Lam\'e constants (or equivalently, $E$ and $\nu$, Young's modulus and Poisson's ratio), and $\mathcal{A}$, $\mathcal{B}$, and $\mathcal{C}$, the (third-order) Landau constants. 
Euler's formula at order $(B/L)^2$, equation (\ref{euler}), involves only one elastic constant, $E$. 
It is thus natural to ask whether Poisson's ratio, $\nu$, plays a role in the non-linear correction to Euler formula  of $(B/L)^4$, the next-order term.
The final answer is found as formula \eqref{correc} below, which shows that the non-linear correction involves all five elastic constants.


\section{Finite compression and buckling}


In this section, we recall the equations governing the homogeneous compression of a cylinder in the theory of exact (finite) elasticity. 
We also recall the form of some incremental solutions that is, of some small-amplitude static deformations which may be superimposed upon the large compression, and which indicate the onset of instability for the cylinder.


\subsection{Large deformation}


We take a cylinder made of a hyperelastic, compressible, isotropic solid with strain-energy density $W$ say, with radius $B$ and length $L$ in its undeformed configuration.
We denote by ($R$, $\Theta$, $Z$) the coordinates of a particle in the cylinder at rest, in a cylindrical system.

Then we consider that the cylinder is subject to the following deformation,
\begin{equation}  \label{deformation}
{r} = \lambda_1R, \qquad {\theta}= \Theta, \qquad {z}=\lambda_3 Z,
\end{equation}
where ($r$, $\theta$, $z$) are the current coordinates of the particle initially at  ($R$, $\Theta$, $Z$),   $\lambda_1$ is the radial stretch ratio and $\lambda_3$ is the axial stretch ratio.
Explicitly, $\lambda_1 = b/B$ and $\lambda_3 = l/L$, where $b$ and $l$ are the radius and length of the deformed cylinder, respectively.
The physical components of  $\tens{F}$, the corresponding deformation gradient, are: $\tens{F} = \textrm{Diag} \left(\lambda_1, \lambda_1, \lambda_3 \right)$, showing that the deformation is equi-biaxial and homogeneous (and thus, universal). 

The (constant) Cauchy stresses required to maintain the large homogeneous compression are 
(see, for instance, Ogden \cite{Ogden84}),
\begin{equation} \label{Cauchy stress tensor}
\sigma_i = \left( \det \tens{F} \right)^{-1}\lambda_i W_i, \qquad i=1,2,3 \quad \textrm{(no sum)},
\end{equation}
where $W_i \equiv \partial W/ \partial \lambda_i$.
In our case, $\sigma_1=\sigma_2$ because the deformation is equi-biaxial, and  $\sigma_1=\sigma_2=0$ because the outer face of the cylinder is free of traction. 
Hence
\begin{equation}
\sigma_1=\sigma_2= \lambda_1^{-1}\lambda_3^{-1}W_1=0, \qquad 
\sigma_3= \lambda_1^{-2} W_3. \label{sigma}
\end{equation}
Note that we may use the first equality to express one principal stretch in terms of the other (provided, of course, that inverses can be performed).


\subsection{Incremental equations}


Now we recall the equations governing the equilibrium of incremental solutions, in the neighborhood of the finite compression.
They read in general as \cite{Ogden84}
\begin{equation} \label{Incremental_equations_equilibrium}
\textrm{div}\ \tens{s}= \mathbf{0},
\end{equation}
where $\tens{s}$ is the incremental nominal stress tensor.
It is related to $\mathbf{u}$, the incremental mechanical displacement, through the incremental constitutive law,
\begin{equation} \label{component_form}
\mathbf{s} = \tens{B} \left(\text{grad} \ \mathbf{u}\right)^T,
\end{equation}
where $\tens{B}$ is the fourth-order tensor of incremental elastic moduli and the gradient is computed in the current cylindrical coordinates \cite{DoHa06}. 
The non-zero components of $\tens{B}$, in a coordinate system aligned with the principal axes associated with the deformation \eqref{deformation}, are given in general by \cite{Ogden84}
\begin{align} 
& JB_{iijj}= \lambda_i\lambda_j W_{ij}, \nonumber \\
&JB_{ijij}= (\lambda_i W_i - \lambda_j W_j)\lambda_i^2/(\lambda_i^2-\lambda_j^2), &  i\neq j,\ \lambda_i\neq \lambda_j, \nonumber \\
&JB_{ijji}= (\lambda_j W_i - \lambda_i W_j)\lambda_i \lambda_j/(\lambda_i^2-\lambda_j^2), &  i\neq j,\ \lambda_i\neq \lambda_j, \nonumber \\
& JB_{ijij}=(B_{iiii}-B_{iijj} + \lambda_i W_i)/2, & i\neq j,\ \lambda_i= \lambda_j, \nonumber \\
& JB_{ijji}=B_{jiij}=B_{ijij} - \lambda_i W_i, &  i\neq j,\ \lambda_i= \lambda_j, 
\end{align}
(no sums), where $W_{ij}\equiv \partial^2 W/(\partial \lambda_i\partial \lambda_j)$. 
Note that here, some of these components are not independent one from another because $\lambda_1=\lambda_2$ and $\sigma_1=\sigma_2=0$. 
In particular, we find that 
\begin{align}
& B_{1212} = B_{2121} = B_{1221}, \qquad 
   B_{2323}=B_{1313}=B_{1331}=B_{2332},
   \notag \\
& B_{2222}=B_{1111}, \quad 
   B_{2233}=B_{1133},
   \quad
    B_{3232}=B_{3131},  \quad 
   B_{1122}=B_{1111}-2B_{1212}. 
\end{align}


\subsection{Incremental solutions}


We look for incremental static solutions that are periodic along the circumferential and axial directions, and have yet unknown radial variations. 
Thus our ansatz for the components of the mechanical displacement is the same as Wilkes's \cite{Wilk55}:
\begin{equation} \label{slon}
u_r = U_r(r) \cos n\theta \cos kz, \quad
u_\theta = U_{\theta}(r) \sin n \theta \cos kz, \quad
u_z = U_z(r) \cos n\theta \sin k z, 
\end{equation}
where $n=0,1,2,\ldots$ is the \textit{circumferential mode number}; $k$ is the \textit{axial wavenumber}; the subscripts $(r,\theta,z)$ refer to components within the cylindrical coordinates $(r,\theta,z)$; and all upper-case functions are functions of $r$ alone. 

Dorfmann and Haughton \cite{DoHa06} show that the following displacements $\mathbf{U}^{(1)}$,  $\mathbf{U}^{(2)}$, and  $\mathbf{U}^{(3)}$ are solutions to the incremental equations \eqref{Incremental_equations_equilibrium}, 
\begin{equation} \label{U12}
\mathbf{U}^{(1)}(r), \, \mathbf{U}^{(2)}(r) = 
\left[ I'_{n}(qkr), -\frac{n}{qkr}I_n(qkr), -\dfrac{(B_{1111}q^2-B_{3131})}{q(B_{1313}+B_{1133})}I_n(qkr)\right]^T,
\end{equation}
and 
\begin{equation} \label{U3}
\mathbf{U}^{(3)}(r) = \left[\dfrac{1}{r}I_n(q_3kr), - \dfrac{q_3k}{n}I'_{n}(q_3kr), 0\right]^{T},
\end{equation}
where $q=q_1, q_2$ and $I_n$ is the modified Bessel function of order $n$.
Here $q_1$, $q_2$, and $q_3$ are the square roots of the roots $q_1^2$, $q_2^2$ of the following quadratic in $q^2$: 
\begin{equation} 
B_{1313} B_{1111} q^4+[(B_{1133}+B_{1313})^2-B_{1313}B_{3131}-B_{3333}B_{1111}]q^2+B_{3333}B_{3131}=0,
\end{equation}
and of the root of the following linear equation in $q^2$
\begin{equation}
B_{1212}q^2-B_{3131}=0,
\end{equation}
respectively.

From \eqref{component_form} we find that the incremental traction on planes normal to the axial direction has components of the same form as that of the displacements, namely 
\begin{align} \label{soln}
& s_{r r} = S_{r r}(r) \cos n\theta \cos k z, \notag \\
& s_{r\theta} = S_{r\theta}(r) \sin n \theta \cos k z, \notag \\
& s_{r z} = S_{r z}(r) \cos n\theta \sin k z, 
\end{align}
say, where again all upper-case functions are functions of $r$ alone. 
Then we find that the traction solutions corresponding to the solutions \eqref{U12}-\eqref{U3} are given by
\begin{multline}
r \mathbf{S}^{(1)}(r), \ \ r \mathbf{S}^{(2)}(r)= 
\left[ 
2B_{1212} I'_n(q k r) \right.
\\ \left . - \left(2B_{1212}  \dfrac{n^2}{q k r} + q k r B_{1111} - \dfrac{k r B_{1133}\left(B_{1111}q^2 - B_{3131}\right)}{q\left(B_{1313}+B_{1133}\right)}\right) I_n(q k r), \right.
 \\ 
\left. 2  n B_{1212}\left(\frac{I_n(q k r)}{q k r}-I'_n(q k r)\right),
 -  k r B_{1313}\left(\frac{B_{1111}q^2-B_{3131}}{B_{1313}+B_{1133}}+1\right)I'_{n}(q k r)
\right]^T,
\end{multline}
and
\begin{multline}
r \mathbf{S}^{(3)}(r) = \left[
 2 B_{1212}\left(\frac{I_n(q_3kr)}{r}-q_3 k I'_n(q_3kr)\right), \right.
 \\
\left. B_{1212}\left(\frac{2q_3k}{n}I'_n(q_3kr)-\left(\frac{2n}{r}+\frac{q_3^2k^2r}{n}\right)I_n(q_3kr)\right),
-  k B_{1313}I_n(q_3kr)\right]^{T}.
\end{multline}

The general solution to the incremental equations of equilibrium is thus of the form
\begin{equation} 
r \mathbf{S}(r)  = 
 \begin{bmatrix} [c|c|c]
r \mathbf{S}^{(1)}(r) & r \mathbf{S}^{(2)}(r) & r \mathbf{S}^{(3)}(r)
\end{bmatrix}
\mathbf{c},
\end{equation}
where $\mathbf{S} \equiv [S_{r r}, S_{r\theta}, S_{r z}]^T$ and $\mathbf{c}$ is a constant three-component vector.
Note that we use the quantity $r \mathbf{S}$ for the traction (instead of $\mathbf{S}$), because it is the Hamiltonian conjugate to the displacement in cylindrical coordinates \cite{Shuv03}.

\begin{figure}[!ht]
\begin{center}
\epsfig{figure=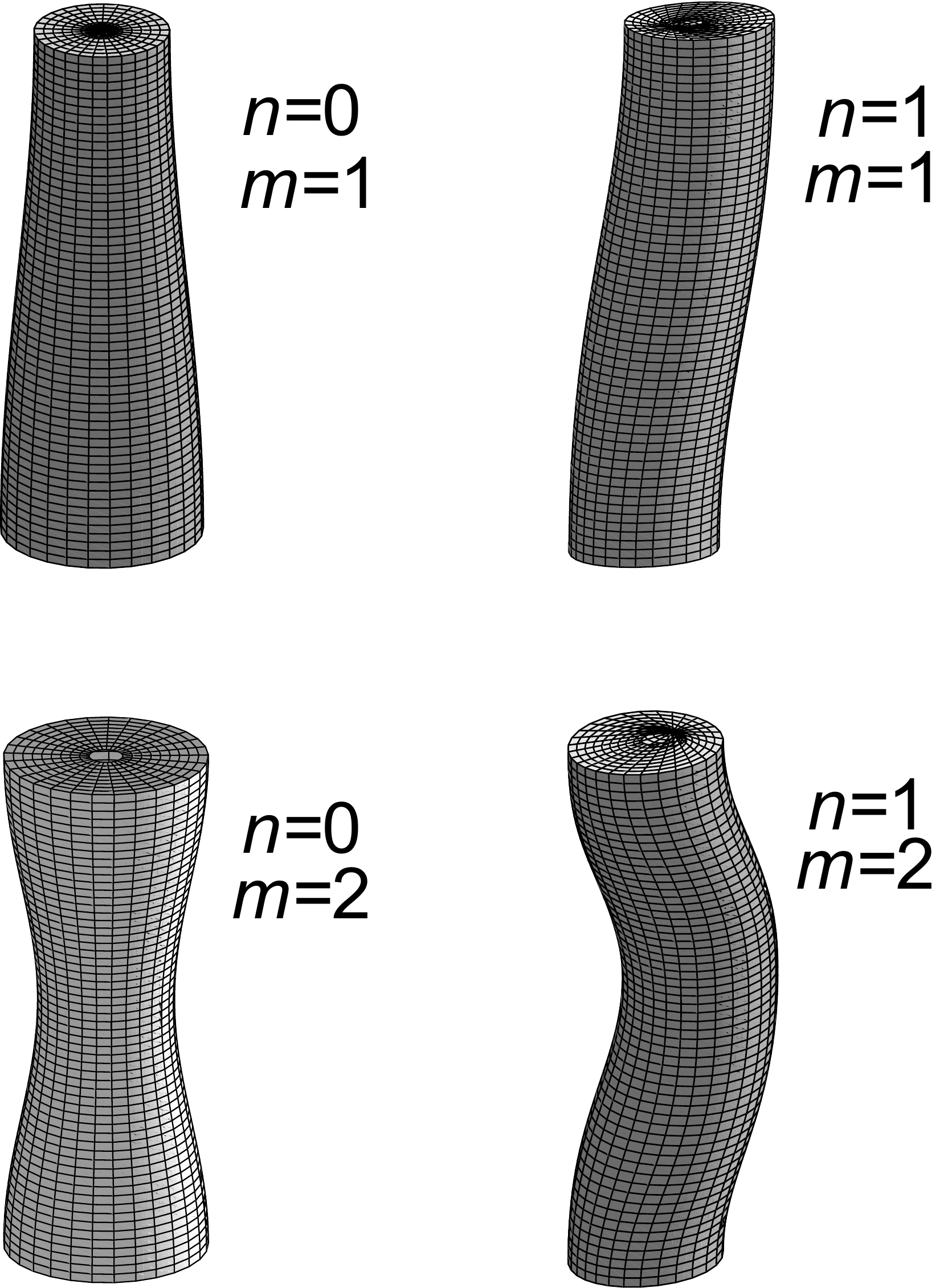, width=.6\textwidth}
\end{center}
 \caption{First two axi-symmetric and two asymmetric modes of buckling for a compressed strut with guided-guided end conditions: $n$ is the circumferential mode number and $m$ the axial mode number. For slender enough cylinders, the $n=1$, $m=1$ mode is the first mode of buckling.}
 \label{fig4}
\end{figure}

Now when the cylinder is compressed (by platens say), its end faces should stay in full contact with the platens so that the first incremental boundary condition is 
\begin{equation}
u_{z} = 0, \qquad \text{on} \quad z=0, l, 
\end{equation}
which leads to \cite{DoHa06,Goriely08}
\begin{equation}
k = m \pi/l,\end{equation}
for some integer $m$, the \emph{axial mode number}.
From \eqref{soln}, we now see that on the thrust faces, we have
\begin{equation}
s_{r z} = 0, \qquad \text{on} \quad z=0, l, 
\end{equation}
which means that the end faces of the column are in sliding contact with the thrusting platens.
In other words, in the limit of a slender column, we recover the Euler strut with \emph{sliding-sliding}, or \emph{guided-guided} end conditions.  
In Figure \ref{fig4}, we show the first two axi-symmetric and two asymmetric modes of incremental buckling.

The other boundary condition is that the cylindrical face is free of incremental traction: $\mathbf{S}(b) = \mathbf{0}$.
This gives 
\begin{equation}
\Delta \equiv \det \begin{bmatrix} [c|c|c]
b \mathbf{S}^{(1)}(b) & b \mathbf{S}^{(2)}(b) & b \mathbf{S}^{(3)}(b)
\end{bmatrix} = 0.
\end{equation}


\section{Euler buckling}




\subsection{Asymptotic expansions}
\label{Asymptotic Euler buckling}


We now specialize the analysis to the asymmetric buckling mode $n=1$, $m=1$, corresponding to the Euler buckling with guided-guided end conditions, in the limit where the axial compressive stretch $\lambda_3$ is close to 1 (the other modes are not reached for slender enough cylinders). 
To this end, we only need to consider the so-called \emph{third-order elasticity} expansion of the strain energy density, for example that of Landau and Lifshitz \cite{LaLi86}, 
\begin{equation}
W = 
\dfrac{\lambda}{2}\left(\textrm{tr}\tens{E}\right)^2 +\mu \ \textrm{tr}(\tens{E}^2) + \dfrac{\mathcal{A}}{3}\textrm{tr}(\tens{E}^3) + \mathcal{B}\left(\textrm{tr}\tens{E}\right)\textrm{tr}(\tens{E}^2)
+  \dfrac{\mathcal{C}}{3}\left(\textrm{tr}  \tens{E}\right)^3,
\end{equation}
where $\tens{E}=\tens{E}^T$ is the Lagrange, or Green, strain tensor defined as $\tens{E}=\left(\tens{F}^T\tens{F}-\tens{I}\right)/2$,
$\lambda$ and $\mu$ are the Lam\'e moduli, and $\mathcal{A}$, $\mathcal{B}$, $\mathcal{C}$ are the Landau third-order elastic constants
(Note that there are other, equivalent, expansions based on other invariants, such as the ones proposed by Murnaghan \cite{Murn51}, Toupin and Bernstein \cite{ToBe61},  Bland \cite{Blan69}, or Eringen and Suhubi \cite{ErSu74}, see Norris \cite{Norr99} for the connections.)

To measure how close $\lambda_3$ is to 1, we introduce $\epsilon$, a small parameter proportional to the slenderness of the  deformed cylinder,
\begin{equation} \label{epsilon}
\epsilon=k b=\pi b/l.
\end{equation}
Then we expand the radial stretch $\lambda_1$ and the critical buckling stretch $\lambda_3$ in terms of $\epsilon$ up to order $M$,
\begin{equation}
\lambda_1=\lambda_1(\epsilon)=1+\sum_{p=1}^{M}\alpha_p\epsilon^p + \mathcal{O}(\epsilon^{M+1}),
\qquad
\lambda_3=\lambda_3(\epsilon)=1+\sum_{p=1}^{M}\beta_p\epsilon^p + \mathcal{O}(\epsilon^{M+1}),
\end{equation}
say, where the $\alpha$'s and $\beta$'s are to be determined shortly. 
Similarly, we expand $\Delta$ in powers of $\epsilon$,
\begin{equation}\nonumber
\Delta = \sum_{p=1}^{M_d}d_p\epsilon^p + \mathcal{O}(\epsilon^{M_d+1}),
\end{equation}
say, and solve each order $d_p=0$ for the coefficients $\alpha_p$ and $\beta_p$, making use of the condition $\sigma_1=0$.
We find that $\alpha_p$ and $\beta_p$ vanish identically for all odd values of $p$, and that $\lambda_1$ and $\lambda_3$, up to the fourth-order in $\epsilon$, are given by 
\begin{equation}\label{lambda1_3}
\lambda_1=1+\alpha_{2}\epsilon^2+\alpha_{4}\epsilon^4+O(\epsilon^6),
\qquad
\lambda_3=1+\beta_{2}\epsilon^2+\beta_{4}\epsilon^4+O(\epsilon^6),
\end{equation}
with $\alpha_2$ and $\alpha_4$ given by 
\begin{align} \label{alpha}
 \alpha_2 = & \dfrac{\nu}{4}, \notag  \\
 \alpha_4 = & -  \dfrac{\nu(1+\nu)}{32} \notag \\ 
&  - \dfrac{(1+\nu)(1-2\nu)}{16E} \left[\nu^2\mathcal{A} + (1-2\nu + 6\nu^2)\mathcal{B} + (1 - 2\nu)^2 \mathcal{C} \right] - \nu \beta_4,  
\end{align}
wherein 
\begin{align} \label{beta}
 \beta_2 = & -\dfrac{1}{4},  \notag \\
 \beta_4  = & \dfrac{29 + 39 \nu + 8 \nu^2}{96(1+\nu)}\notag \\
 & + \dfrac{1}{16E} \left[( 1 - 2\nu^3)\mathcal{A} + 3(1-2\nu)(1+2\nu^2)\mathcal{B} + (1 - 2\nu)^3 \mathcal{C} \right]. 
\end{align}
Note that we switched from Lam\'e constants to Poisson's ratio and Young's modulus for these expressions, using the  connections $\nu = \lambda/(2\lambda+2\mu)$ and $E = \mu(3\lambda+2\mu)/(\lambda+\mu)$.
%


\subsection{Onset of nonlinear Euler buckling}


The analytical results presented above are formulated in terms of the \emph{current} geometrical parameter $\epsilon$, defined in \eqref{epsilon}. 
In order to relate these results to the classical form of Euler buckling, we introduce the \emph{initial} geometric slenderness $B/L$. 
Recalling that $\epsilon=\pi b/l$, $\lambda_3=l/L$, and $b=\lambda_1B$, we find that 
\begin{equation}\label{condition}
\epsilon \lambda_3 = \pi  \lambda_1 (B/L).
\end{equation} 
We expand $\epsilon$ in powers of $B/L$, and solve \eqref{condition} to obtain 
\begin{eqnarray}\label{epsilon2}
\nonumber \epsilon 
&=& \pi (B/L)  + (\alpha_2-\beta_2) \pi^3 (B/L)^3 
+ \mathcal{O}\left((B/L)^4\right) \\
&=& \pi (B/L)  + (1 + \nu)(\pi^3/4)(B/L)^3 + \mathcal{O}\left((B/L)^4\right).
\end{eqnarray}

Second, we wish to relate the axial compression to the current axial load $N$. 
To do so, we integrate the axial stress over the faces of the cylinder, 
\begin{equation} \label{axialforce}
 N = -2\pi\int_0^b r \sigma_3 \text{d}r = -\pi b^2 \sigma_3=-\pi \lambda_1^2 B^2 \sigma_3,
\end{equation}
because $\sigma_3$ is constant, given by (\ref{sigma})$_{2}$.

Finally, in order to write the nonlinear buckling formula, we expand $\lambda_1$ and $\lambda_3$ in  \eqref{axialforce}, using \eqref{lambda1_3}, and then expand $\epsilon$ in powers of the slenderness ($B/L$), using \eqref{epsilon2}.
It gives the desired expression for the first \emph{non-linear correction to the Euler formula},
\begin{equation} \label{correc}
\dfrac{N_\text{cr}}{\pi^3 B^2} = \dfrac{E}{4}  \left(\dfrac{B}{L}\right)^2 - \dfrac{\pi^2}{96} \delta_\text{\ NL} \left(\dfrac{B}{L}\right)^4,
\end{equation}
where
\begin{multline}
\delta_\text{\ NL} = 2\dfrac{13 + 12 \nu - 2 \nu^2}{(1+\nu)} E \\
 + 12\left[(1 - 2\nu^3)\mathcal{A} + 3 (1 - 2\nu)(1 + 2\nu^2)\mathcal{B} + (1 - 2\nu)^3\mathcal{C}\right]. 
\end{multline}

We now check this equation against its incompressible counterpart \eqref{incompressible_case}.
Theoretical considerations and experimental measurements \cite{Ogden74b}, \cite{WHIZ08}, \cite{CaGF03}, \cite{DeOg10}, show that in the incompressible limit, $E$ and $\mathcal{A}$ remain finite, $\nu \rightarrow 1/2$, $(1-2\nu)\mathcal{B} \rightarrow  -E/3$, and $(1-2\nu)^3 \mathcal{C} \rightarrow 0$.
It is then a simple exercise to verify that \eqref{correc} is indeed consistent with \eqref{incompressible_case} in those limits.


\subsection{Examples}

\begin{figure}[!ht]
\centering \mbox{\subfigure{\epsfig{figure=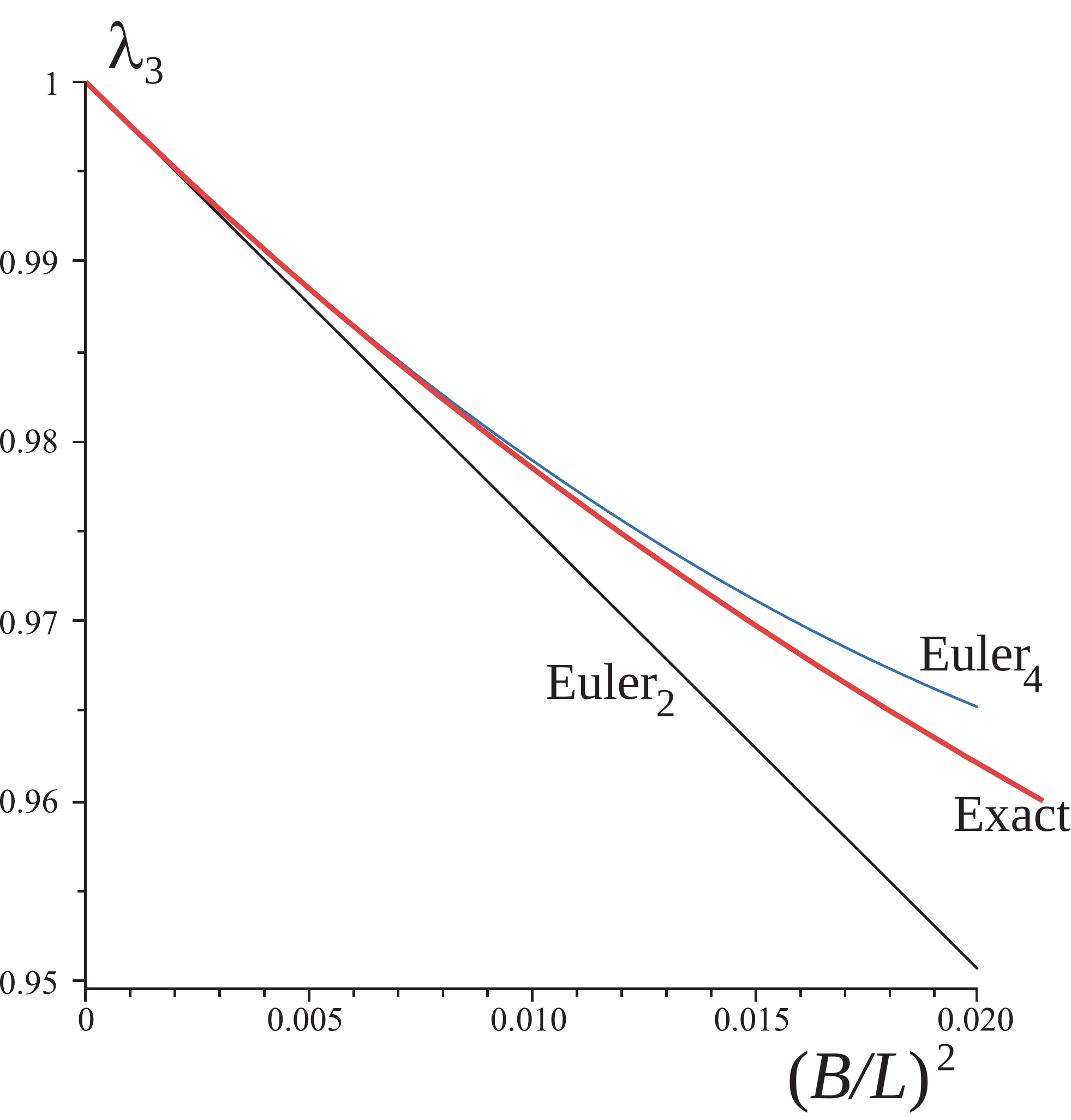, width=.45\textwidth}}}
  \quad \quad
     \subfigure{\epsfig{figure=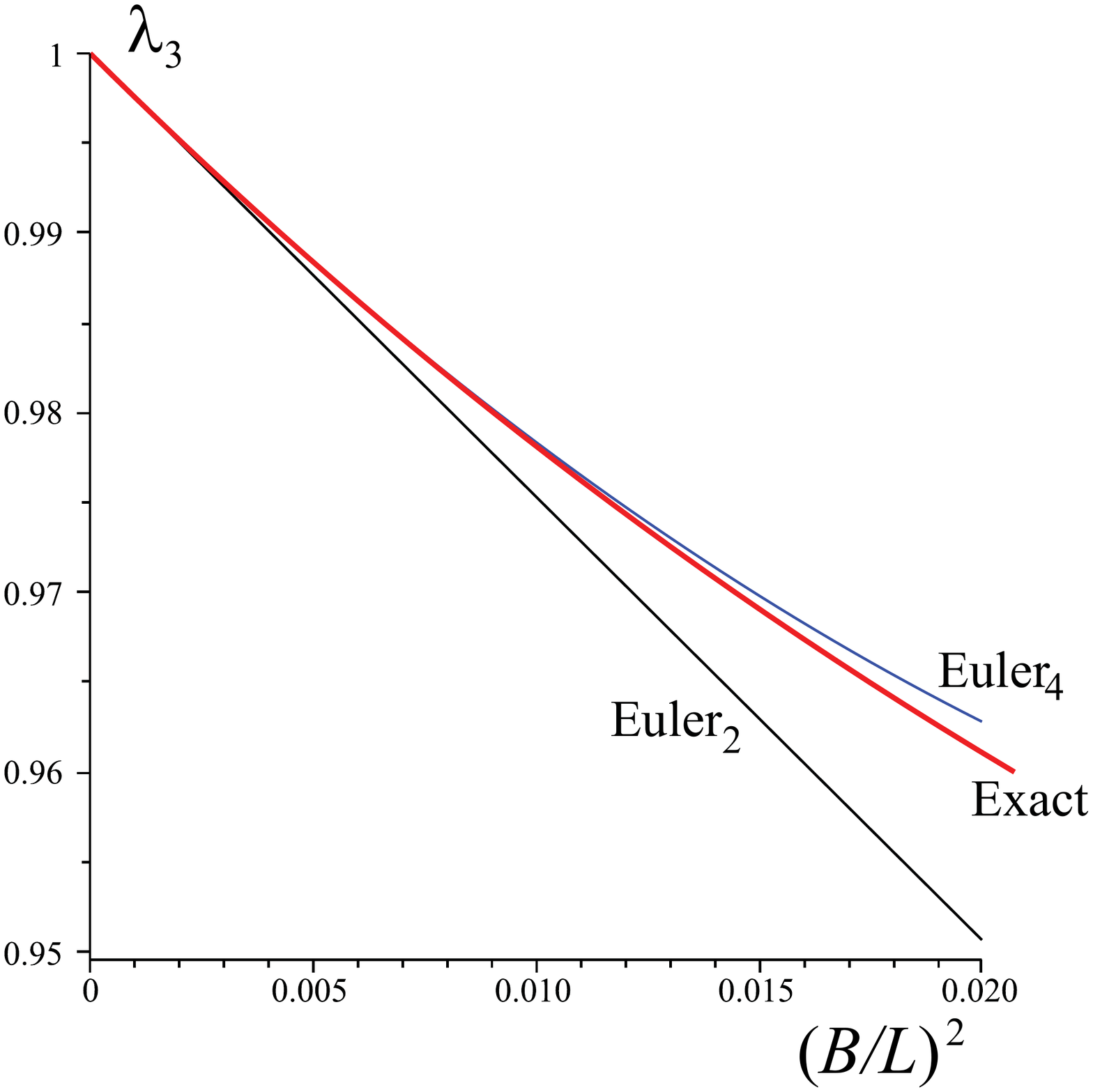, width=.45\textwidth}}
 \caption{Comparison of the different Euler formulas obtained by expanding the exact solution to order 2 (classical Euler buckling formula, plot labeled ``Euler$_2$'') and to order 4 (plot labeled ``Euler$_4$''), for pyrex (figure on the left) and for silica (figure on the right).}
 \label{fig5}
\end{figure}

\begin{table}[!ht]
\caption{Lam\'e constants and Landau third-order elastic moduli for five solids ($10^{9}$ N$\cdot$ m$^{-2}$)}
\label{tab:1}      
\begin{tabular}{lrrrrr}
\hline\noalign{\smallskip}
material & $\lambda$ & $\mu$ & $\mathcal{A}$ & $\mathcal{B}$ & $\mathcal{C}$ \\
\noalign{\smallskip}\hline\noalign{\smallskip}
      Polystyrene & $ 1.71$ & $0.95$ & $-10$ & $-8.3$ & $-10.6$ \\ 
			Steel Hecla 37 & $111$ & $82.1$ & $-358$ & $-282$ & $-177$ \\
			Aluminium 2S  & $57$ & $27.6$ & $-228$ & $-197$ & $-102$ \\
			Pyrex glass & $13.5$ & $27.5$ & $420$ & $-118$ & $132$ \\
			SiO$_2$ melted & $15.9$ & $31.3$ & $-44$ & $93$ & $36$ \\ 
			\noalign{\smallskip}\hline
\end{tabular}
\end{table}
To evaluate the importance of the non-linear correction, we computed the critical axial stretch ratio of column buckling for two solids.
In Table \ref{tab:1}, we list the second- and third-order elastic constants of five compressible solids, as collected by Porubov \cite{Poru03} (in the Table we converted the ``Murnaghan constants'' given by Porubov to the Landau constants $\mathcal{A,B,C}$).
Figure \ref{fig5} shows the variations of $\lambda_3$ with the squared slenderness $(B/L)^2$, for pyrex and silica (two last lines of Table 1).


\section{Conclusions}


The present analysis provides an asymptotic formula for the critical value of the load for the Euler buckling problem, with guided-guided (sliding-sliding) end conditions. This formula was checked both in the incompressible limit and on particular cases against the exact value of the buckling obtained from the exact solutions. Not surprisingly it reinforces the universal and generic nature of the Euler buckling formula as the correction is small for most systems even when nonlinear elastic effects and nonlinear geometric effects are taken into account. It would be of great interest to see if these effects could be observed experimentally.




\begin{thebibliography}{}

\bibitem{tige61}
Timoshenko, S. P., Gere, J. M.:
Theory of elastic stability
Mc graw-Hill, New York (1961)


\bibitem{Goriely08}
Goriely, A., Vandiver, R., Destrade, M.: 
Nonlinear Euler buckling. 
Proc. Roy. Soc. A, 464, 3003-3019 (2008). 

\bibitem{Wilk55}
Wilkes, E.W.:
On the stability of a circular tube under end thrust.
Quart. J. Mech. appl. Math., 8, 88-100 (1955).

\bibitem{Biot63}
Biot, M.A.:
Surface instability of rubber in compression.
Appl. Sc. Research A, 12, 168-182 (1963).

\bibitem{FoSh63}
Fosdick, R.A., Shield, R.T.:
Small bending of a circular bar superposed on finite extension or compression.
Arch. Rational Mech. Analysis, 12, 223-248 (1963).

\bibitem{Ogden74b}
Ogden, R.W.: 
On isotropic tensors and elastic moduli. 
Proc. Camb. Phil. Soc., 75, 427-436 (1974). 

\bibitem{Ogden84}
Ogden, R.W.: 
Non-Linear Elastic Deformations. 
Dover, New York (1984).

\bibitem{DoHa06}
Dorfmann, A., Haughton, D.M.: 
Stability and bifurcation of compressed elastic cylindrical tubes.
Int. J. Eng. Sc., 44, 1353-1365 (2006).

\bibitem{Shuv03}
Shuvalov, A.L.:
A sextic formalism for three-dimensional elastodynamics of cylindrically anisotropic radially inhomogeneous materials.
Proc. Roy. Soc. A, 459, 1611-1639 (2003).

\bibitem{LaLi86}
Landau, L.D., Lifshitz, E.M.:
Theory of Elasticity, 3rd ed. Pergamon, New York, 1986.

\bibitem{Murn51}
Murnaghan, F.D.: 
Finite Deformations of an Elastic Solid. 
Wiley, New York, 1951.

\bibitem{ToBe61}
Toupin, R.A., Bernstein, B.: 
Sound waves in deformed perfectly elastic materials. Acoustoelastic effect. 
J. Acoust. Soc. Am. 33, 216�225 (1961).

\bibitem{Blan69}
Bland, D.R.: 
Nonlinear Dynamic Elasticity. 
Blaisdell, Waltham (1969).

\bibitem{ErSu74}
Eringen, A.C., Suhubi, E.S.:  
Elastodynamics, Vol. 1. 
Academic Press, New York (1974).

\bibitem{Norr99}
Norris, A. N.: 
Finite amplitude waves in solids, In: M. F. Hamilton and
D. T. Blackstock (eds.) 
Nonlinear Acoustics.
Academic Press, San Diego, pp. 263�277 (1999).

\bibitem{Poru03}
Porubov, A.V.: 
Amplification of Nonlinear Strain Waves in Solids.
World Scientific, Singapore (2003).

\bibitem{WHIZ08}
Wochner, M.S., Hamilton, M.F., Ilinskii, Y.A., Zabolotskaya, E.A.: 
Cubic nonlinearity in shear wave beams with different polarizations.
J. Acoust. Soc. Am., 123, 2488-2495 (2008).

\bibitem{CaGF03}
Catheline, S., Gennisson,  J.-L., Fink, M.:
Measurement of elastic nonlinearity of soft solid with transient elastography.
J. Acoust. Soc. Am., 114, 3087-3091 (2003).

\bibitem{DeOg10}
Destrade, M., Ogden, R.W.:
On the third- and fourth-order constants of incompressible isotropic elasticity.
submitted.


\end{thebibliography}
\end{document}